\documentclass[12pt]{article}
\usepackage{graphics}
\begin{document}
\title{New physics in $t\overline t$ spin correlations at the Tevatron}
\author{\normalsize Bob Holdom\thanks{bob.holdom@utoronto.ca}~~{\small\it and}~~Tibor
Torma\thanks{kakukk@physics.utoronto.ca}\\[1ex]
  \small\em Department of Physics, University of Toronto\\
  \small\em Toronto, Ontario, Canada M5S\ 1A7}
\maketitle
\begin{picture}(0,0)
\put(310,205){UTPT-99-06}
\end{picture}
\begin{abstract}We show that the angular distributions of leptons or jets due to $t\overline t$
spin correlations allow a determination of the top chromomagnetic moment $\kappa$ with an accuracy
of order 0.1. The method is quite insensitive to background, event reconstruction, and other
experimental uncertainties. The total event number is important, so we suggest the inclusion of all-hadronic
events.
\end{abstract}
%
\baselineskip 19pt

\section{Introduction}

A primary goal of high energy experiments is to observe deviations from
the standard model. Prior to the detection of, or in the absence of, new light resonances
such as a Higgs particle, one is left with low energy remnants of loop effects, such as
higher dimensional operators. The dimension-5 top chromomagnetic moment operator is of 
special interest due to its relationship with the top mass\footnote{Our signs and normalization are such
that the gauge coupling is $- g_s G^a_\mu \  \overline t\,\frac{\lambda^a}{2}\gamma^\mu t$. The
convention for the sign of
$\kappa$ is the same (opposite) to the first (second) paper in Ref.~\cite{Rizzo}. }
\begin{equation}
- g_s \frac{\kappa}{4m_t}\, G^a_{\mu\nu} \  \overline t\,\frac{\lambda^a}{2}\sigma^{\mu\nu} t.
\end{equation}
In dynamical schemes of the top mass the estimate is $\kappa \approx m_t^2/\Lambda^2$, where
$\Lambda$ characterizes some scale of physics from which the top is deriving its mass. This is typically the
electroweak symmetry breaking scale or smaller. The assumption here is that the heavier degrees of freedom
also carry color, in which case the chromomagnetic moment is related to the top mass diagram by attaching a
gluon. The situation is different for the standard model or other models involving elementary scalar fields. In
these models the chromomagnetic moment arises at one loop, in contrast to the tree level origin
of the top mass, and this leads to a $1/(4\pi)^2$ loop factor suppression of the
chromomagnetic moment.

The effect of a top chromomagnetic moment in hadron collider experiments has been
investigated ~\cite{Rizzo}. Naively, one would expect that the energy dependence of the
new contribution differs significantly from the standard model contribution, as the derivative acting on the
top field introduces a momentum factor. Actually the structure of the contributions is schematically like
$1+\kappa+\kappa^2{q^2\over m_t^2}$. The would-be enhancement factor $q\over m_t$ in the
$\kappa$ term is absent, since the interference between the standard chirality conserving amplitude and the
chirality changing ($\kappa$) amplitude requires an additional $m_t$ factor. This, together with the fact
that most $t\bar t$ pairs are produced not far above the threshold, explains the fact ~\cite{Rizzo} that
various differential cross sections in top momentum  variables are sensitive to $\kappa$ essentially only
through an overall change in their normalization. The extraction of $\kappa$ from such observables at the
Tevatron is thus difficult because (i) the luminosity is not known with a good accuracy, and 
(ii) other operators may also contribute to the structureless excess.

We are led consider other observables. In particular a chirality
changing contribution should affect the spin correlations of the $t\overline t$
state, thus modifying the angular distributions of the top decay products.
Analyses of the $t\overline t$ spin structure in the dominant $q\overline q\to g\to t\overline t$
process in the standard model are provided
in~Refs.~\cite{MahlonParke1,Will,ParkShadmi,MahlonParke3}. Some of the top spin information resides
in the direction of the charged lepton (or down-type quark), the decay product of the top in the chain $t\to
W\to l$. When only this direction is measured in the top rest frame, relative to the top spin direction $s$, the
spin correlation shows up as~\cite{MahlonParke3,MahlonTalk}
\begin{equation}
\frac{1}{(d\sigma/d\cos\theta)}\,\frac{d\sigma}{d\cos\theta\,d\cos\vartheta\,d\cos\overline\vartheta}=
\frac{1}{4}\left[1+\frac{N_{\times}-N_{\parallel}}{N_{\times}+N_{\parallel}}
\, \cos\vartheta \cos\overline\vartheta\right].\label{eq:mp}
\end{equation}
$\vartheta$ is the angle between $s$ and the charged lepton direction, and similarly for
$\overline\vartheta$ and $\overline s$ on the
$\overline t$ side. ($\theta$ is the top scattering angle.) $N_\parallel$ ($N_\times$) is the
number like-spin (unlike-spin)
$t\overline t$ pairs (by like-spins we mean spins both parallel or both antiparallel to their respective axes). 
$N_\parallel$ vanishes for a particular choice of $s$, which defines the spin basis dubbed
``off-diagonal"~\cite{MahlonParke3}. Then the $t$ and $\overline t$ spins are $100\%$ correlated,
up to small corrections from the gluon induced process.

Unfortunately the $\vartheta$ and $\overline\vartheta$ variables as defined in the off-diagonal basis
provide a poor probe of the chromomagnetic moment operator. We have
calculated the differential cross section for $q\overline q\to g\to t\overline t\to b\,U\overline D+\overline
b\,\overline UD$ to ${\cal O}(\kappa)$ in terms of an enlarged set of phase space variables. (At Tevatron
energies
${\cal O}(\kappa^2)$ effects may safely be ignored.) Requiring that both the top quarks and the $W^\pm$
be on shell we find, again with variables defined in the off-diagonal basis,
\begin{eqnarray}\label{eq:zhat}
&&\hspace{-0.9in}
\frac{d\sigma}
{d\cos\theta\,d\cos\vartheta\,d\varphi\,d\cos\overline\vartheta\,d\overline\varphi}\propto{\cal R}_\kappa
\equiv{\cal R}^{(0)}+{\cal R}^{(1)}\kappa
\\
&=&
(1+\cos\vartheta\cos\overline\vartheta)
-\frac{\beta^2}{2}\sin^2\theta\left[1+\cos\vartheta\cos\overline\vartheta-
\sin\vartheta\sin\overline\vartheta\cos(\varphi-\overline\varphi)\right]
\nonumber\\
&+&\kappa\left\{
2 (1+\cos\vartheta\cos\overline\vartheta)-
\gamma\beta^2\sin\theta\cos\theta\left(\cos\vartheta\sin\overline\vartheta\sin\overline\varphi+
\cos\overline\vartheta\sin\vartheta\sin\varphi\right)\right\}.\nonumber
\end{eqnarray}
$\varphi$ and $\overline\varphi$ are the azimuthal
angles around the corresponding spin axes $s$ and $\overline s$ and 
$\gamma=\left(1-\beta^2\right)^{-1/2}$ where $\beta$ is the top velocity.

We immediately observe that an integration over the azimuthal variables gives back
\begin{equation}\label{eq:fact}
\frac{d\sigma}{d\cos\theta\,d\cos\vartheta\,d\cos\overline\vartheta}\propto
\left(1+2\kappa-\frac{\beta^2}{2}\sin^2\theta\right)
\left(1+\cos\vartheta\cos\overline\vartheta\right).
\end{equation}
This result clearly elucidates the
relationship of our approach to both Ref.~\cite{Rizzo} and~\cite{MahlonParke3}. The first factor contains
all information used in~\cite{Rizzo}, with the $1/2$ factor suppressing the dependence on both $\beta$ and
$\theta$. The second factor is the same as
found in~\cite{MahlonParke3} (Eq.~(\ref{eq:mp}) above in the off-diagonal basis), so that the $\vartheta$
and $\overline\vartheta$ variables by themselves introduce no dependence on~$\kappa$.

The complicated structure of the cross section formula prompts us to look for a simple variable which still
carries most of the information useful for the extraction of $\kappa$ from a data sample. Attempts to find
such variables were discussed in~\cite{Kingman}. But the structure of Eq.~(\ref{eq:zhat}) is such 
that all useful information is encoded in the
$\sin$'s and $\cos$'s of the azimuthal angles which change sign as we move around the phase space. 
The best variable should incorporate as much of this information as possible. 

Although not necessarily simple, this ``best'' variable is provided by the maximum likelihood method, which
is the standard way of extracting the value of a parameter of a distribution from a sample. Noting that ${\cal
R}_\kappa$ (Eq.~(\ref{eq:zhat})) is defined to be linear in
$\kappa$, this variable is
\begin{equation}\label{eq:F}
F=\frac{{\cal R}^{(1)}}{{\cal R}^{(0)}}.
\end{equation}
The maximum likelihood method uses only the values of this variable in the determination of $\kappa$.

In order to assess the observability of $\kappa$ in Run~2 of the Tevatron we built a Monte-Carlo generator
and used a maximum likelihood estimate on the generated data sample to determine the possible statistical
accuracy of the estimated value of $\kappa$. We found, taking a reasonable $1500$ events, a
\mbox{``1-sigma"} accuracy of ${\cal O}(\pm0.1)$. What is quite remarkable is that the accuracy of the
determination is very stable against factors like (i) the introduction of various cuts, (ii) crude inaccuracies in
the measurement of the parameters of each individual jet, (iii) the inability to tell apart the two hadronic
decay products of the $W$ or (iv) the introduction of a background even as large as ten times the 
$t\overline t$ signal. All these insensitivities are due to the complicated but smoothly varying function
in Eq.~(\ref{eq:F}), which is not easily faked or influenced by any of the above factors.

The only factor of any importance is the number of $t\overline t$ events. We find that a sensible
determination of $\kappa$ requires $1000$-$2000$ events and any increase in the event number
significantly improves the determination. However, for each individual event all that is needed is
a very crude estimate of the directions of all six decay products (on the partonic level), which
is even possible in the case of all-hadronic decay. It seems advantageous to relax the standard set of cuts, at
the cost of larger backgrounds and loss of precision in each event, in order to gain in the event number. The
method may then provide an estimate of $\kappa$ superior to the determination from the total
cross section. In any case, the method advocated here gives a completely independent way to help determine
the source of any observed cross section anomaly.

The rest of the paper is organized as follows. Sec.~\ref{sec:calcul} describes our calculation
of the spin density matrix, explains how we handle the six-particle phase
space, and produces our result for the differential cross section. Sec.~\ref{sec:ML} describes the maximum
likelihood estimate method while gives Sec.~\ref{sec:stab} gives the numerical results.

\section{The structure of the differential cross section}\label{sec:calcul}

\begin{figure}[htb]
\begin{center}
\includegraphics{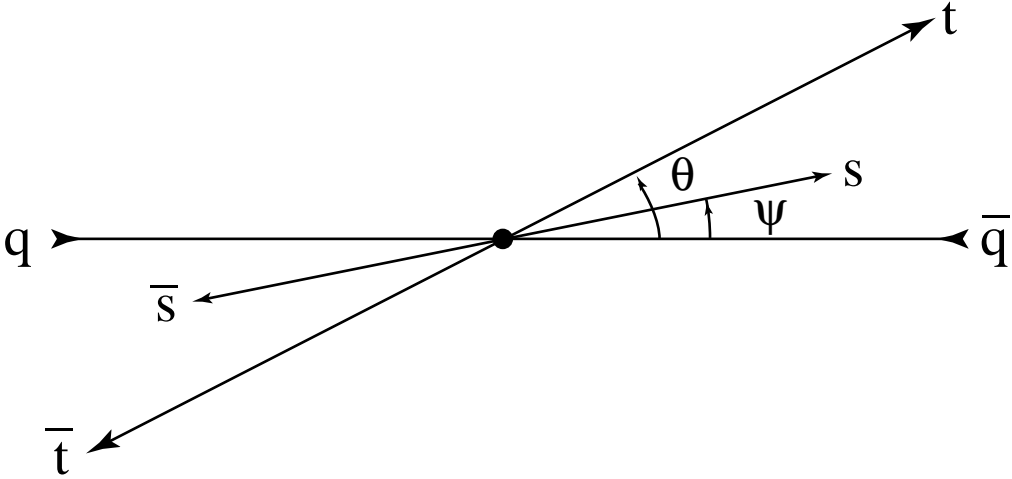}
\end{center}
\caption{\protect\label{fig:spinbasis}The definition of a spin basis in the ZMF frame.
The $z$-axis points out of the page and is used to define the azimuthal angles $\varphi$
and $\overline\varphi$.}
\end{figure}

At the Tevatron the dominating process for $t\bar t$ production is quark fusion
$q\overline q\to g(k)\to t\overline t$. The amplitude of this process at tree level is
\begin{equation}\label{eq:topamp}
{\cal M}_{t\overline t}=
\frac{i}{4}\,\frac{g_s^2}{\hat s}\left(\lambda^a_{q\overline q}\lambda^a_{t\overline t}\right)
\left(\overline v_{\overline q}\gamma_\mu u_q\right)\left[\overline u_t
\left(\gamma^\mu+\frac{i}{2}\,\frac{\kappa}{m_t}\,\sigma^{\mu\nu}k_\nu\right)v_{\overline t}\right].
\end{equation}
which directly translates into a spin density matrix for the top
\begin{equation}\label{eq:defrho}
\sum|{\cal M}_{t\overline t}|^2=\frac{16\pi^2\alpha_s^2}{9}
\rho_{s\overline s,s^\prime\overline s^\prime},
\end{equation}
where color summation/averaging is understood, as well as averaging for incoming quark spins.
The indices $s,\overline s$ refer to the top (antitop) spin, and $s^\prime,\overline s^\prime$
are the corresponding spin indices in the conjugate amplitude. They are all measured with respect
to an arbitrary (for the moment) spin direction $s^\mu,\overline s^\mu$. For definiteness, we
choose the spatial part of the $s^\mu,\overline s^\mu$ vectors back-to-back at an angle~$\psi$
to the beam direction in the zero momentum frame (ZMF) frame (see Fig.~\ref{fig:spinbasis}).

We calculate each of the elements of the density matrix. Its structure is greatly simplified
in the ``off-diagonal" basis defined by~\cite{MahlonParke3}
\begin{equation}
\tan\psi=\frac{\beta^2 s_\theta c_\theta}{1-\beta^2s^2_\theta}\ \ \mbox{~with~~}
-\frac{\pi}{2}<\psi<\frac{\pi}{2}.
\end{equation}
In this basis the standard model contributions from like-spin $t\overline t$ pairs vanish. We determine how
the result in~\cite{MahlonParke3} is modified at linear order in $\kappa$:
\begin{eqnarray}\label{eq:offdiagrho}
\rho&\longrightarrow& \left[\begin{array}{cccc}
\uparrow\uparrow,\uparrow\uparrow & \uparrow\uparrow,\uparrow\downarrow & \uparrow\uparrow,\downarrow\uparrow & \uparrow\uparrow,\downarrow\downarrow \\
\uparrow\downarrow,\uparrow\uparrow & \uparrow\downarrow,\uparrow\downarrow & \uparrow\downarrow,\downarrow\uparrow & \uparrow\downarrow,\downarrow\downarrow \\
\downarrow\uparrow,\uparrow\uparrow & \downarrow\uparrow,\uparrow\downarrow & \downarrow\uparrow,\downarrow\uparrow & \downarrow\uparrow,\downarrow\downarrow \\
\downarrow\downarrow,\uparrow\uparrow & \downarrow\downarrow,\uparrow\downarrow & \downarrow\downarrow,\downarrow\uparrow & \downarrow\downarrow,\downarrow\downarrow 
\end{array}\right]_{s\overline s,s^\prime\overline s^\prime} \\
&\equiv&\left[\begin{array}{cccc}
0&i\kappa\,\gamma\,\beta^2 s_\theta c_\theta&-i\kappa\,\gamma\,\beta^2 s_\theta c_\theta&0\\
-i\kappa\,\gamma\,\beta^2 s_\theta c_\theta&2(1+2\kappa)-\beta^2s^2_\theta&\beta^2s^2_\theta&i\kappa\,\gamma\,\beta^2 s_\theta c_\theta\\
i\kappa\,\gamma\,\beta^2 s_\theta c_\theta&\beta^2s^2_\theta&2(1+2\kappa)-\beta^2s^2_\theta&-i\kappa\,\gamma\,\beta^2 s_\theta c_\theta\\
0&-i\kappa\,\gamma\,\beta^2 s_\theta c_\theta&i\kappa\,\gamma\,\beta^2 s_\theta c_\theta&0
\end{array}\right]\nonumber
\end{eqnarray}
We use the shorthands $s_\theta=\sin\theta$ and $c_\theta=\cos\theta$. The $\uparrow$ states
are those where the $t$ ($\overline t$) have spin projection  $+1/2$ on their
respective spin axes. Note that this matrix is rank~2, as in the standard model case, so it involves only two
$t\overline t$ spin states. Any other choice of the angle $\psi$ (such as $\psi=0$ for the
helicity basis) only makes the form of $\rho$ more complicated.

All the $\kappa$-dependence of the process is encoded in this density matrix. Note that
$\mbox{Tr}\,\rho=4(1+2\kappa-\frac{1}{2}\beta^2s^2_\theta)$ is 
nothing but the first factor in~Eq.~(\ref{eq:fact}). The information in the off-diagonal elements
will be observed in the angular distributions of the decay products.
\begin{figure}[tbh!]
\begin{center}
\includegraphics{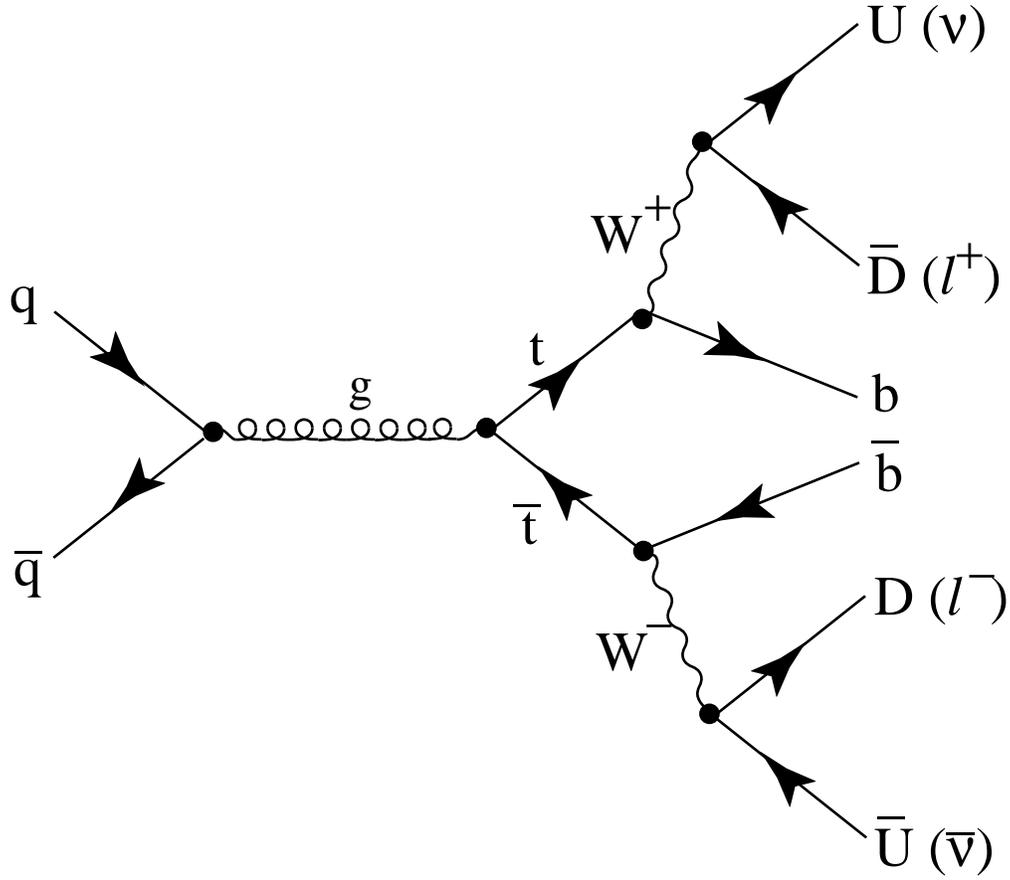}
\end{center}
\caption{\protect\label{fig:6part}The full cascade of $t\overline t$ production and decay. The decay
products of the $W^+$ are an up-type $U=u,c$ quark (or a neutrino) and a down-type antiquark
$\overline D=\overline d, \overline s$ (or a positively charged lepton, $e^+$ or $\mu^+$).}
\end{figure}

The correct choice of the variables in the six particle phase space (see Fig.~\ref{fig:6part})
should keep the expressions for both the amplitude and the six-particle phase space measure as
simple as possible. We should also make sure that the regions of integration are simple. For example,
no combination of angles and/or energies in the ZMF frame will satisfy these requirements. For these
reasons we have chosen the following set of variables, in addition to the ones that define the ZMF frame:
\begin{itemize}
\item{The scattering angle $\theta$ between $q$ and $t$ in ZMF and the corresponding (irrelevant)
azimuthal angle defining the scattering plane.}
\item{The angle $\vartheta$ in the top frame between the spin direction $s$ and the $\overline D$ quark
(or charged lepton) and the corresponding azimuthal angle $\varphi$. Note that
$\varphi=0$ corresponds to the positive $z$ direction in~Fig.~\ref{fig:spinbasis}.}
\item{The angle $\overline\vartheta$ in the antitop frame between the spin direction $\overline s$ and the
$D$ quark (or charged lepton) and the corresponding azimuthal angle $\overline\varphi$. Note that 
$\overline\varphi=0$ also corresponds to the same positive $z$
direction in~Fig.~\ref{fig:spinbasis}.}
\item{The angle $\lambda$ in the top frame between the direction of $W^+$ and the $\overline D$ quark and the
corresponding azimuthal angle $\phi$; also the respective angles $\overline\lambda,\overline\phi$
on the $\overline t$ side.}
\end{itemize}

These ten variables completely describe the six-particle phase space (which is 10 dimensional due
to four restrictions from momentum conservation and four from the on-shell conditions of the
$t,\overline t, W^\pm$). Their corresponding ranges are completely independent of each other
and the phase space measure is very simple. We emphasize though that
these angles are not all defined in the same frame.

We calculated the decay rate of each of the 16 spin states of the $t\overline t$
system individually and used a Breit-Wigner formula to find the $q\overline q\to6\,\mbox{quarks}$
cross section.
\begin{eqnarray}\label{eq:sigma}
\sigma&=&\frac{m_W^4m_t^6(m_t^2-m_W^2)^2}{(m_t^2+m_W^2)^6\Gamma_t^2\Gamma_W^2}
\,\frac{\alpha_s^2\alpha_W^4}{9(2\pi)^32^{14}}
\,|V_{tb}^2V_{U\overline D}V_{\overline UD}|^2\,\\
&&\times\left(\int_{-1}^1d\cos\lambda\oint d\phi\frac{1-\cos\lambda}{(1-\beta_W\cos\lambda)^4}\right)
\left(\int_{-1}^1d\cos\overline\lambda\oint d\overline\phi\frac{1-\cos\overline\lambda}
{(1-\beta_W\cos\overline\lambda)^4}\right)\nonumber\\
&&\times \,\frac{\beta}{\gamma^2}\ \int_{-1}^1d \cos\theta \int_{-1}^1d\cos\vartheta \oint d\varphi 
\int_{-1}^1d\cos\overline\vartheta \oint d\overline\varphi\ \ {\cal R}_\kappa
\left(\theta,\vartheta,\varphi,\overline\vartheta,\overline\varphi\right).\nonumber
\end{eqnarray}
Here $\beta_W=({m_t^2-m_W^2})/({m_t^2+m_W^2})$ and we
note how the integration over the four
variables $\lambda,\phi,\overline\lambda,\overline\phi$ can be trivially done. We find
\begin{equation}\label{eq:diffrate}
{\cal R}_\kappa=\frac{1}{4}\,\sum_{s\overline s\,s^\prime\overline s^\prime}
\rho_{s\overline s,s^\prime \overline s^\prime}(\theta)
\left[\begin{array}{cc}
1+\cos\vartheta&\sin\vartheta e^{i\varphi}\\
\sin\vartheta e^{-i\varphi}&1-\cos\vartheta
\end{array}\right]_{ss^\prime}
\left[\begin{array}{cc}
1-\cos\overline\vartheta&\sin\overline\vartheta e^{i\overline\varphi}\\
\sin\overline\vartheta e^{-i\overline\varphi}&1+\cos\overline\vartheta
\end{array}\right]_{\overline s\,\overline s^\prime},
\end{equation}
where the two matrices represent the angular dependences of the $t$ and $\overline t$
decay rates. The structure of the above matrices implies that the off-diagonal elements of
$\rho$ gets encoded in the $\varphi$ and $\overline\varphi$-distributions. By integrating
out $\varphi$ and $\overline\varphi$ we would see only the diagonal elements of $\rho$.
Substituting Eq.~(\ref{eq:offdiagrho}) into Eq.~(\ref{eq:diffrate}) gives the result for~${\cal R}_\kappa$
in Eq.~(\ref{eq:zhat}).
Thus we have managed to express all spin correlation information in terms of five angles and
the top quark velocity.

\section{Maximum likelihood and the best variable}\label{sec:ML}

We would like now to assess the statistical errors in the estimation of~$\kappa$
at Run~2 of the Tevatron. For that purpose we simulate the actual process by a sample of events generated
by a ``theorist's Monte-Carlo", an event generator which uses the quark-level differential cross sections
in~Eqs.~(\ref{eq:zhat},\ref{eq:sigma}) folded with MRST parton distribution functions. For our purposes
of crude estimation it is sufficient to treat gluon induced processes as background and to ignore higher loop
and radiative corrections.

As the next step we took the generated sample with a given small \mbox{$\kappa=\kappa_{true}$} and used an
unbinned maximum likelihood method to estimate $\kappa$ from the sample. We calculate the
log-likelihood function
\begin{equation}\label{eq:loglp}
\log{\cal L}(\kappa)=\sum_i\log p_\kappa^{(i)}
\end{equation}
where $p_\kappa^{(i)}$ is the probability for an event to occur at the point of the phase space
labeled by $i$. This probability obviously depends on the measure used in the phase space but
one observes that all the ambiguity in redefining the phase space variables (and so, the measure) and
in redefining the grid used to produce a nonzero probability for each event produces only
a $\kappa$-independent Jacobian factor multiplying ${\cal L}(\kappa)$. We remove this ambiguity
altogether by subtracting the standard model value $\log{\cal L}(\kappa=0)$ from Eq.~(\ref{eq:loglp}).
This will not affect the shape of the likelihood function nor the position of its maximum.

The quantity $p_\kappa$, however, involves a nontrivial normalization issue. As a probability it should be
normalized to one:
\begin{equation}\label{eq:pullout}
p_\kappa^{(i)}=\frac{f(\Phi_i)+\kappa\,f^\prime(\Phi_i)}
{\int d\Phi f(\Phi)+\kappa\, \int d\Phi f^\prime(\Phi)}
\equiv
\frac{f(\Phi_i)}{\int d\Phi f(\Phi)}\times\frac{1+\kappa\,\frac{f^\prime(\Phi_i)}{f(\Phi_i)}}
{1+\kappa\, \frac{\int d\Phi f^\prime(\Phi)}{\int d\Phi f(\Phi)}}.
\end{equation}
Recall that we are considering the amplitude-squared to linear order in $\kappa$. It is a
function of the phase space variables collectively denoted as $\Phi$.
The first factor in Eq.~(\ref{eq:pullout}), being independent of~$\kappa$, can be absorbed into the
normalization ambiguity of the likelihood function, while the last denominator cannot. It is the reflection
of the fact that we estimate $\kappa$ on a sample of given size, so we loose the information related
to the total rate. On the other hand we are keeping the information residing in the angular distributions,
and in this way our approach is orthogonal to that in~\cite{Rizzo}.

The probability of each event in Eq.~(\ref{eq:pullout}) is expressed in terms of one variable
$F(\Phi)=\frac{f^\prime(\Phi)}{f(\Phi)}$ and a constant
$\Delta=\frac{\int d\Phi f^\prime(\Phi)}{\int d\Phi f(\Phi)}$, so
\begin{equation}\label{eq:logl}
\log{\cal L}(\kappa)=\sum_i\log \frac{1+\kappa\,F_i}{1+\kappa\,\Delta}.
\end{equation}
The variable $F$ carries all information used by the maximum likelihood method and is thus
the ``best" variable appearing in~Eq.~(\ref{eq:F}).

\section{The stability of the estimation}\label{sec:stab}

We estimated~$\kappa$ in samples of various sizes, with
various true values of $\kappa$ fed into the Monte-Carlo generator. As can be understood from
Eq.~(\ref{eq:logl}), the likelihood function becomes singular at $\kappa=-F_i^{-1}$ for each event $i$.
These singularities are an artifact of our linear approximation in $\kappa$, and tell us that our
estimation is valid only when $\kappa\ll\mbox{min}\,|F_i^{-1}|$. But this turns out to cover the
theoretically interesting range of~$\kappa$.
\begin{figure}[htb!]
\begin{center}
\includegraphics{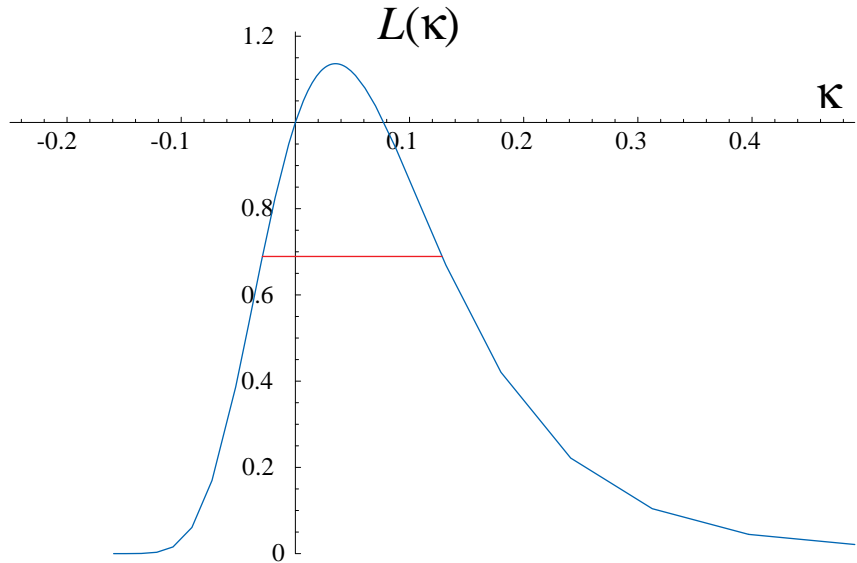}
\end{center}
\caption{\protect\label{fig:Lfunct}
The likelihood function for a specific Monte-Carlo run with
\mbox{$\kappa_{true}=0$}, based on 1500 events.
The horizontal line represents the \mbox{1-sigma} interval. The skewing of the function is common
to the different runs, but the location of the maximum is of course randomly distributed about
$\kappa=0$.}
\end{figure}

Fig.~\ref{fig:Lfunct} shows the likelihood function for a characteristic example run
with 1500 observed $t\overline t$ events.
Following tradition, we will quote the interval where the log-likelihood function drops by $1/2$
as a \mbox{1-sigma} interval, although we have to keep in mind that such a statement is mathematically
inaccurate. This interval is shown as the horizontal line. We see that the estimation
in~Fig.~\ref{fig:Lfunct} is consistent with the true value of~$\kappa$ and corresponds to a $85\%$
probability level.

We ran the Monte-Carlo with various different initial settings of the random
generator. We found the following \mbox{1-sigma} intervals by averaging the values:
\begin{equation}\label{eq:nocuts}
\begin{array}{ll}
N=200&\Delta\kappa=\left\{{\begin{array}{c}+0.30\\-0.15\end{array}}\right.\\[1em]
N=900&\Delta\kappa=\left\{{\begin{array}{c}+0.22\\-0.12\end{array}}\right.\\[1em]
N=1500&\Delta\kappa=\left\{{\begin{array}{c}+0.10\\-0.07\end{array}}\right.\\[1em]
N=2000&\Delta\kappa=\left\{{\begin{array}{c}+0.10\\-0.07\end{array}}\right.\\[1em]
N=5000&\Delta\kappa=\left\{{\begin{array}{c}+0.05\\-0.03\end{array}}\right.
\end{array}
\end{equation}

We immediately observe the importance of the event number. The number of observed $t\bar t$ events
on Run~2 is expected to be ${\cal O}(1500)$. With that number the accuracy is in the order of
$\Delta\kappa=0.1$, not very far from the prediction of some models of new physics.

We next address the question whether experimental realities significantly deteriorate these
results. We considered the following issues:
\begin{itemize}
\item{In the hadronic decay mode of the $W$ it is almost impossible to tell which of the two jets
comes from which quark. We set up a similar maximum likelihood estimation of $\kappa$ where the new
probability function now accounts for a $50\%$ probability of misidentifying the two jets. The change
from the values in~(\ref{eq:nocuts}) is insignificant (by this we will mean a less than $20\%$ change in $\Delta\kappa$.)
This is true for ``symmetrization" on only one side (i.e.~for the $W^+$ only, relevant for
semileptonic events) and for both sides, relevant for all-hadronic decays.
\item{We imposed several sets of cuts on the data to see their effect on the extraction. The variable
$F_i$ is unmodified but the normalization constant 
$\Delta=\frac{\int d\Phi f^\prime(\Phi)}{\int d\Phi f(\Phi)}$ is modified as the integration region
is changed. We found a value for $\Delta$ by estimating $\kappa$ on a very large sample
and requiring that it reproduce $\kappa=\kappa_{true}$. For example, we imposed
the same parton level cuts that had been used in the analysis of the semileptonic events in
Run~1~\cite{leptocuts}, (a charged lepton with $E_T>8\,GeV$, $|\eta|<1.0$, a neutrino with
$E_T>20\,GeV$, three jets with $E_T>15\,GeV$ and $|\eta|<2.0$ and one more jet with $E_T>8\,GeV$
and $|\eta|<2.4$. We found only insignificant changes in the accuracy when the number of events passing
the cuts were fixed. Imposing a jet separation
cut of $\Delta\alpha>20^0$ similarly had little effect.}
\item{There are usually very large uncertainties in extracting the quark three momenta. We set up
a random distortion of the energy and direction of each jet by a relative portion of $\delta=30\%$.
Imposing the top and $W$ mass conditions on the data we found again only insignificant changes. As a check,
$\delta=90\%$ did destroy the usefulness of the data as would be expected.}
\item{Finally, we included a very crude model of the background, for which we used a uniform
distribution in the angular variables. For the number of $t\overline t$ events constant,
$N=1500$, the size of the \mbox{1-sigma} interval increases as follows:
\begin{equation}\label{eq:bkg}
\begin{array}{lll}
N_{total}=15000&\Delta\kappa=\left\{\begin{array}{c}+0.15\\-0.09\end{array}\right.\\[1em]
 N_{total}=4500&\Delta\kappa=\left\{\begin{array}{c}+0.13\\-0.08\end{array}\right.\\[1em]
N_{total}=3000&\Delta\kappa=\left\{\begin{array}{c}+0.11\\-0.07\end{array}\right.\\[1em]
N_{total}=1875&\Delta\kappa=\left\{\begin{array}{c}+0.11\\-0.07\end{array}\right.\\[1em]
N_{total}=1500&\Delta\kappa=\left\{\begin{array}{c}+0.10\\-0.07\end{array}\right.
\end{array}
\end{equation}
This slight increase, however, shows only that even a large background can be tolerated in
exchange for a larger event number.}}
\end{itemize}

We see that many of the experimental difficulties have little effect in our case, 
basically due to the fact they cannot easily mimic the intricacies of the angular distribution
in~Eq.~(\ref{eq:F}). Systematic errors would similarly be dangerous only if they include a component
with a correlation between the various angles similar to~Eq.~(\ref{eq:F}).

In conclusion, we have shown how $t\overline t$ spin correlation information can be extracted from the
angular distributions of the $t\overline t$ decay products. This provides
an independent and competitive way to observe new physics contributions to the chromomagnetic moment
of the top quark. The statistical limitation on the extraction of~$\kappa$ at Run~2 of the Tevatron is ${\cal
O}(\pm0.1)$, mainly coming from statistics. The extraction is surprisingly stable against experimental
uncertainties and we suggest that cuts should be relaxed as much as possible in order to gain in the event
number. The inclusion of all-hadronic events should also be given serious consideration.

\section*{Acknowledgments}

We would like to thank Andrew Robinson for useful discussions. This work was
supported in part by the Natural Sciences and Engineering Research Council of Canada.

\end{document}